\documentclass[
    ,final            
  ]
  {aipproc}
\usepackage{graphicx}

\layoutstyle{6x9}

\begin{document}

\title{ALICE diffractive physics in p-p and Pb-Pb collisions at the LHC}


\author{R. Schicker}{
  address={Phys. Inst., Philosophenweg 12, 69120 Heidelberg}
}


\begin{abstract}
The ALICE experiment at the Large Hadron Collider LHC is presently
being commissioned. ALICE consists of a central barrel, a muon spectrometer
and neutron calorimeters at $0^0$. Additional detectors for event
classification and trigger purposes are located on both sides of 
the central barrel. 
The geometry of the ALICE detector allows the implementation of a 
diffractive double gap trigger by requiring two or more tracks in the 
central barrel but no activity in the event classification detectors.  
Some selected diffractive physics channels are discussed which 
become accessible by a double gap trigger. The interest of such 
diffractive measurements in proton-proton as well as in lead-lead
collisions is outlined. 

\end{abstract}

\maketitle


\section{The ALICE Experiment}

The ALICE experiment is designed for taking data in the high
multiplicity environment of lead-lead collisions at the Large Hadron 
Collider (LHC)\cite{Alice1,Alice2}. The ALICE experiment 
consists of a central barrel covering the pseudorapidity 
range $-0.9 < \eta < 0.9$ and a muon spectrometer in the 
range $-4.0<\eta<-2.4$. Additional detectors for trigger purposes and 
for event classification exist in the range $ -4.0 < \eta < 5.0 $.

\subsection{The ALICE Central Barrel}

The detectors in the ALICE central barrel track and identify 
hadrons, electrons and photons in the pseudorapidity range 
$ -0.9 < \eta < 0.9$. The magnetic field strength 
of \mbox{0.5 T} allows the measurement of tracks from very low transverse 
momenta  of about \mbox{100 MeV/c} to fairly high values of about 100 GeV/c. 
The tracking detectors are designed to reconstruct secondary vertices 
resulting from decays of hyperons, D and B mesons. 
The main detector systems for these tasks are the Inner Tracking
System, the Time Projection Chamber, the Transition Radiation Detector
and the Time of Flight array. These systems cover the full azimuthal
angle within the  pseudorapidity range $ -0.9 < \eta < 0.9$. 
Additional detectors with partial coverage 
of the central barrel are a PHOton Spectrometer (PHOS), an
electromagnetic calorimeter (EMCAL) and  a High-Momentum Particle 
Identification Detector (HMPID). 
   
\subsection{The ALICE Zero Degree Neutron Calorimeter}

The Zero Degree Neutron Calorimeters (ZDC) are placed on both sides of the 
interaction point at a distance of 116 m\cite{ZDC}. The ZDC information 
can be used to select different diffractive topologies. Events of the
types  $pp \rightarrow ppX, pN^{*}X, N^{*}N^{*}X$ will have no signal, 
signal in one or in both of the ZDC calorimeters, respectively. 
Here, X denotes a centrally produced diffractive state from which the 
diffractive L0 trigger is derived. 

\section{The ALICE diffractive gap trigger}

Additional detectors for event classification and trigger purposes
are located on both sides of the ALICE central barrel. First, an array
of scintillator detectors (V0) is placed on both sides of the 
central barrel. These arrays are labeled V0A and V0C on the 
two sides, respectively. Each of these arrays covers a pseudorapidity
interval of about two units with a fourfold segmentation of half a 
unit. The azimuthal coverage is divided into eight 
segments of 45$^{0}$ degrees hence each array is composed  of 32
individual counters.  
Second, a Forward Multiplicity Detector (FMD) is located on both sides 
of the central barrel. The pseudorapidity coverage of this detector
is $-3.4 < \eta < -1.7$ and $1.7 < \eta < 5.1$, respectively.

\begin{figure}[htb]    
\includegraphics[height=.28\textheight]{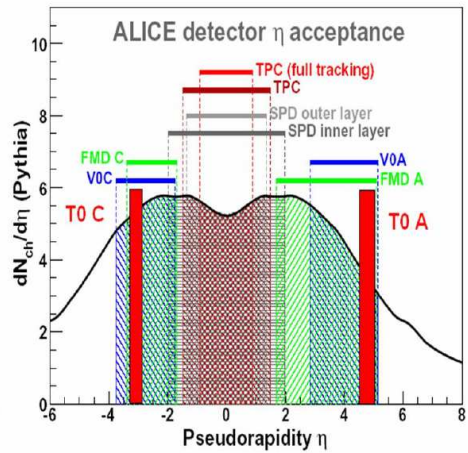}
\caption{Pseudorapidity coverage of trigger detectors and of detectors
in central barrel}
\label{fig:acc_eta}
\end{figure}

Fig.\ref{fig:acc_eta} shows the pseudorapidity coverage of the detector
systems described above. The geometry of the central barrel 
in conjunction with the additional detectors V0 and FMD is well suited
for the definition of a rapidity gap trigger.
Such a gap trigger can be defined by the requirement of signals 
coming from the central barrel detectors while V0 and FMD not showing 
any activity. This scheme requires a trigger signal from within 
the central barrel for L0 decision. The pixel detector of the Inner Tracking
System or the Time Of Flight array can deliver such a signal\cite{pixel}. 

The high level trigger HLT has access to the information of all the 
detectors shown in Fig.\ref{fig:acc_eta} and will hence be able to select
events with rapidity gaps in the range $-4 < \eta < -1$ and 
$1 < \eta < 5$. These gaps extend over seven units of pseudorapidity
and are hence expected to suppress minimum bias inelastic events
by many orders of magnitude.   

In addition to the scheme described above, the ALICE diffractive L0 
trigger signal can be generated from the Neutron ZDC if no central state 
is present in the reaction. A L0 signal from ZDC does not arrive at 
the central trigger processor  within the standard L0
time window. A L0 trigger from ZDC is, however, possible during 
special data taking runs for  which the standard L0 time limit is extended.

\section{ALICE diffractive physics}

The tracking capabilities at very low transverse momenta in
conjunction with the excellent particle identification make ALICE an 
unique facility at LHC to pursue a long term physics program of
diffractive physics. The low luminosity of ALICE
as compared to the other LHC experiments restricts the ALICE physics 
program to reactions with cross section at a level of a few nb per unit 
of rapidity.

\begin{figure}[htb]
\includegraphics*[height=0.25\textheight]{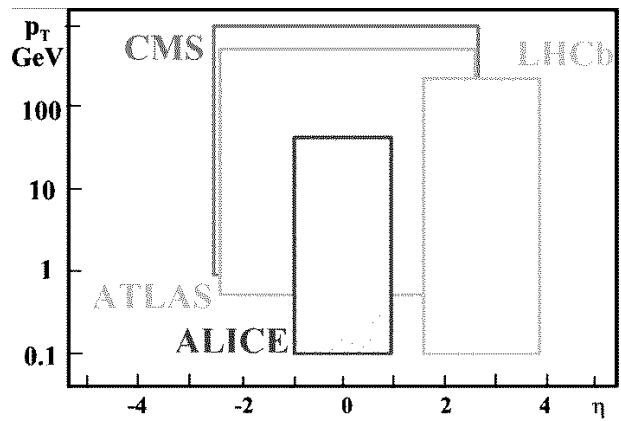}
\caption{Rapidity and transverse momentum acceptance of the LHC experiments}
\label{fig:acc_all}
\end{figure}

Fig.\ref{fig:acc_all} shows the transverse momentum acceptance of the 
four main LHC experiments. Not shown in this
figure is the TOTEM experiment. The acceptance of the
TOTEM telescopes is in the range of $ 3.1 <  | \eta |  < 4.7$ and 
$5.3 < | \eta | <6.5$. 
The CMS transverse momentum acceptance of about 1 GeV/c shown
in Fig.\ref{fig:acc_all} represents a nominal value. The CMS analysis 
framework foresees the reconstruction of a few selected data samples to
values as low as 0.2 GeV/c\cite{CMS}.

\section{Pomeron signatures in proton-proton collisions}

\begin{figure}[htb]
\includegraphics*[height=0.16\textheight]{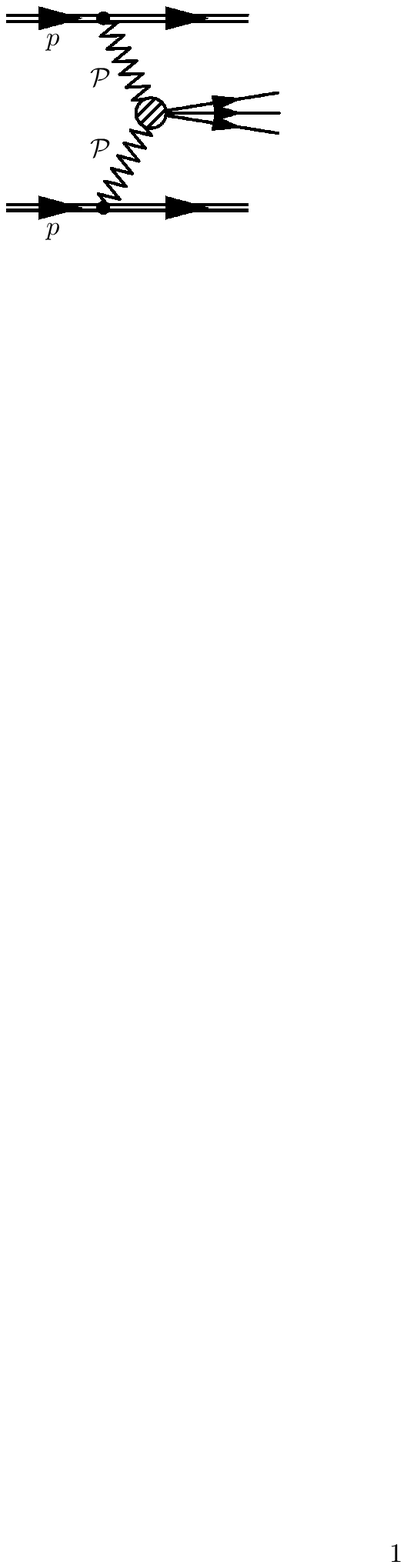}
\caption{Pomeron-pomeron fusion in proton-proton}
\label{fig:ex_pp}
\end{figure}

The ALICE experiment will take data in proton-proton mode at a luminosity of 
\mbox{$\cal{L}$ = $5 x 10^{30} $cm$^{-2}$s$^{-1}$.} 
Double pomeron events in proton-proton collisions as shown in 
Fig.\ref{fig:ex_pp} 
are expected to possess a few interesting properties.  

\begin{itemize}

\item The production cross section of glueball states is expected to
be enhanced in Pomeron fusion events as compared to minimum bias
inelastic events. It will therefore be interesting to study the 
resonances produced in the central region when two rapidity gaps 
are required\cite{close}.   

\item The slope $\alpha'$ of the Pomeron trajectory is rather small:
$\alpha' \sim$ 0.25 GeV$^{-2}$ in DL fit and  $\alpha' \sim$ 0.1
GeV$^{-2}$ in vector meson production at HERA\cite{DL}. These values of
$\alpha'$ in conjunction with the small $t$-slope ($<$ 1 GeV$^{-2}$ ) 
of the triple Pomeron vertex indicate that the mean transverse
momentum $k_t$ in the Pomeron wave function is relatively large 
$\alpha' \sim$ 1/$k_t^2$, most probably $k_t >$ 1 GeV. The transverse 
momenta of secondaries produced in Pomeron-Pomeron interactions are of 
the order of this $k_t$. Thus the mean transverse momenta of secondaries 
produced in Pomeron-Pomeron fusion is expected to be larger as compared to 
inelastic minimum bias events. 

\item The large $k_t$ described above corresponds to a large 
effective temperature. A suppression of strange quark production is 
not expected. Hence the K/$\pi$ ratio is expected to be enhanced in 
Pomeron-Pomeron fusion as compared to inelastic minimum bias 
events\cite{akesson}. Similarly, the $\eta$/$\pi$ and
$\eta'$/$\pi$ ratios are 
expected to be enhanced due to the hidden strangeness content and due 
to the gluon components in the Fock states of $\eta,\eta'$.

\end{itemize}

\section{Pomeron signatures in lead-lead collisions}

ALICE will take data in lead-lead mode at a luminosity of 
$\cal{L}$ = $5 x 10^{27}$cm$^{-2}$s$^{-1}$. 

\begin{figure}[htb]
\includegraphics*[height=0.16\textheight]{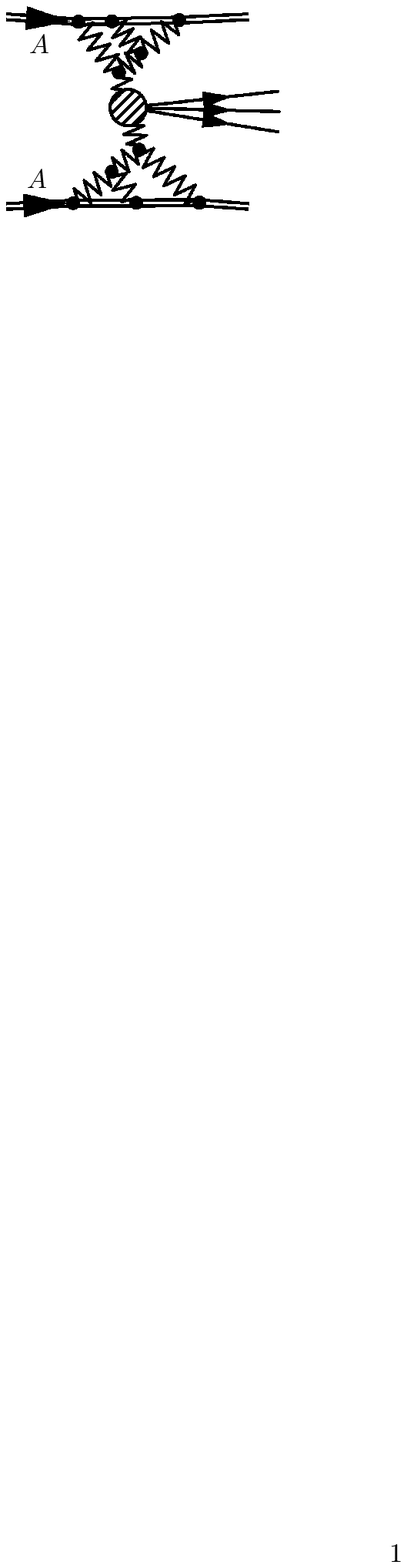}
\caption{Pomeron-pomeron fusion in lead-lead}
\label{fig:ex_AA}
\end{figure}

The cross section of double pomeron induced reaction channels will be 
modified as compared to the proton-proton case due to
absorption and screening as illustrated in \mbox{Fig.\ref{fig:ex_AA}.}
The A-dependence of the cross section for specific pomeron induced
channels hence reflects the contribution of these multi-pomeron diagrams.
The study of multi-pomeron couplings is an important ingredient 
in the analysis of soft diffraction data and in the evaluation of the 
total pp cross section at LHC energies\cite{amartin}.   

\section{Odderon signatures}

Odderon signatures can be looked for in exclusive reactions where
the Odderon (besides the Photon) is the only possible exchange. 
Diffractively produced C-odd states such as vector mesons 
$\phi, J/\psi, \Upsilon$ can result from Photon-Pomeron or
Odderon-Pomeron exchange.  Any excess beyond the Photon contribution
is indication of Odderon exchange. 

Cross section estimates for diffractively produced $J/\psi$ in pp
collisions at LHC energies were first given by Sch\"{a}fer\cite{schaefer}. 
More refined  calculations result in a $t$-integrated photon and
Odderon contribution of $\frac{d\sigma}{dy}\mid_{y=0} \;\sim$ 15 nb 
and 1 nb, respectively\cite{bzdak}. 

If the diffractively produced final state is not an eigenstate of
C-parity, then interference effects between photon-Pomeron and
photon-Odderon amplitudes can be analyzed.

\begin{figure}[htb]
\includegraphics*[height=0.146\textheight]{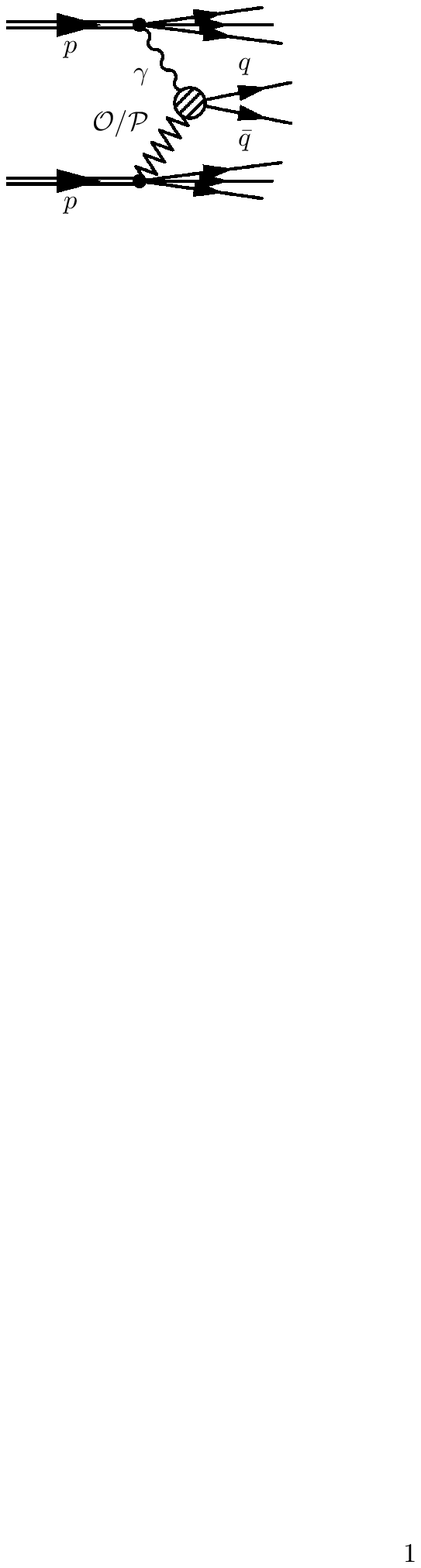}
\caption{photon-Pomeron and photon-Odderon amplitudes}
\label{fig:odderon_inter}
\end{figure}

Fig.\ref{fig:odderon_inter} shows the photon-Pomeron and the
photon-Odderon amplitudes for $q\bar{q}$ production. A study of 
open charm diffractive photoproduction estimates the asymmetry in 
fractional energy to be on the order of 15\%\cite{brodsky}. The 
forward-backward charge asymmetry in diffractive production of pion
pairs is calculated to be on the order of  10\% for pair masses 
in the range 
\mbox{$1\: $GeV/c$^{2} < m_{\pi+\pi-} < 1.3\: $GeV/c$^{2}$\cite{haegler,ginzburg}.}

\begin{theacknowledgments}

I thank Otto Nachtmann and Carlo Ewerz for inspiring conversations.\newline
This work is supported in part by German BMBF under project 06HD197D.
       
\end{theacknowledgments}



\bibliographystyle{aipproc}   

\bibliography{sample}

\IfFileExists{\jobname.bbl}{}
 {\typeout{}
  \typeout{******************************************}
  \typeout{** Please run "bibtex \jobname" to optain}
  \typeout{** the bibliography and then re-run LaTeX}
  \typeout{** twice to fix the references!}
  \typeout{******************************************}
  \typeout{}
 }

\end{document}